\documentclass[traditabstract]{aa}

\usepackage{graphicx}
\usepackage{natbib}
\usepackage{float}
\usepackage{placeins}
\usepackage{caption}
\usepackage[absolute, showboxes]{textpos}
\usepackage{txfonts}
\usepackage{color}

\newcommand{\inten}{erg cm$^{-2}$ s$^{-1}$~}

\newcommand{\cmt}{cm$^{-3}$~}

\newcommand{\gm}{$\Gamma_{\rm mech}$~}

\newcommand{\zsun}{$Z_\odot$~}

\newcommand{\thco}{$^{13}$CO~}
\newcommand{\nhtwo}{$N({\rm H}_2)$~}
\newcommand{\Kkms}{${\rm K~km~s}^{-1}$~}
\newcommand{\kms}{${\rm km~s}^{-1}$~}
\newcommand{\Msun}{${\rm M}_{\odot}$}
\newcommand{\xfco}{X_{\rm CO}}
\newcommand{\xfcomw}{X_{\rm CO, MW}}
\newcommand{\wco}{W_{{\rm CO}(1 - 0)}}
\newcommand{\hcop}{${\rm HCO}^+$~}

\begin{document}
\title{Modelling Mechanical Heating in Star-Forming Galaxies: CO and \thco Line
Ratios as Sensitive Probes}
   \subtitle{}

   \author{M. V. Kazandjian\inst{1}, 
     I. Pelupessy\inst{1}, 
     R. Meijerink\inst{1,2}, 
     F. P. Israel\inst{1}, 
     M. Spaans\inst{2}}

   \institute{Leiden Observatory, Leiden University, P.O. Box 9513, 2300 RA
   Leiden, The Netherlands\\
     \email{mher@strw.leidenuniv.nl}
       \and Kapteyn Astronomical Institute, PO Box 800, 9700 AV Groningen, The
       Netherlands
     }

   \date{Received February 13, 2014; accepted June 6, 2016}

\abstract
{
  We apply photo-dissociation region (PDR) molecular line emission models, that
  have varying degrees of enhanced mechanical heating rates, to the gaseous
  component of simulations of star-forming galaxies taken from the literature.
  Snapshots of these simulations are used to produce line emission maps for the
  rotational transitions of the CO molecule and its \thco~ isotope up to $J = 4-3$.

  We use these maps to investigate the occurrence and effect of mechanical
  feedback on the physical parameters obtained from molecular line intensity
  ratios.

  We consider two galaxy models: a small disk galaxy of solar metallicity and a
  lighter dwarf galaxy with 0.2 \zsun metallicity.  Elevated excitation
  temperatures for CO($1 - 0$) correlate positively with mechanical feedback,
  that is enhanced towards the central region of both model galaxies.  The
  emission maps of these model galaxies are used to compute line ratios of CO
  and \thco~ transitions.  These line ratios are used as diagnostics where we
  attempt to match them   These line ratios are used as diagnostics where we
  attempt to match them to mechanically heated single component (i.e. uniform
  density, Far-UV flux, visual extinction and velocity gradient) equilibrium
  PDR models.
  We find that PDRs ignoring mechanical feedback in the heating budget
  over-estimate the gas density by a factor of 100 and the far-UV flux by
  factors of $\sim 10 - 1000$.
  In contrast, PDRs that take mechanical feedback into account are able to fit
  all the line ratios for the central $< 2$ kpc of the fiducial disk galaxy
  quite well.  The mean mechanical heating rate per H atom that we recover from
  the line ratio fits of this region varies between $10^{-27}$ -- $10^{-26}$~erg
  s$^{-1}$. Moreover, the mean gas density, mechanical heating rate,
  and the $A_V$ are recovered to less than half dex.  On the other hand, our
  single component PDR model fit is not suitable for determining the actual gas
  parameters of the dwarf galaxy, although the quality of the fit line ratios
  are comparable to that of the disk galaxy.
}
\keywords{Galaxies:ISM -- (ISM:) photon-dominated region (PDR) -- ISM:molecules
--
  Physical data and processes:Turbulence}

\authorrunning{Kazandjian {\it et. al}}
\titlerunning{CO and \thco Emission Map Construction for Simulated Galaxies}
\maketitle

\section{Introduction}

Most of the molecular gas in the universe is in the form of H$_2$.  However,
this simple molecule has no electric dipole moment.  The rotational lines
associated with its quadrupole moments are too weak to be observed at gas
temperatures less than 100~K, where star formation is initiated inside clouds
of gas and dust.  This is also true for the vibrational and electronic emission
of H$_2$; hence it is hard to detect directly in the infrared and the far-infrared
spectrum. CO is the second most abundant molecule after H$_2$, and it has been
detected ubiquitously.  CO forms in shielded and cold regions where H$_2$ is present.
Despite its relatively low abundance, it has been widely used as a tracer of
molecular gas.  \cite{solomon75} were the first to establish a relationship
between CO$(1 - 0)$ integrated intensity ($W_{\rm CO(1-0)}$) and H$_2$ column density ($N({\rm
H}_2$)).  Since then, this relationship has been widely used and it is currently
known as the so-called $X$-factor.  The applicability and limitations of the
$X$-factor are discussed in a recent review by \cite{bolatto2013-1}.

Environments where cool H$_2$ is present allow the existence of CO, and many
other molecular species.  In such regions, collisions of these molecules with
H$_2$ excite their various transitions, which emit at different frequencies. 
The emission line intensities can be used to understand the underlying physical
phenomena in these regions. The line emission can be modeled by solving for
the radiative transfer in the gas.  One of the most direct ways to model the
emission is the application of the large velocity gradient (LVG) approximation
\citep{sobolev1960}.
LVG models model the physical state of the gas such as the density and
temperature but do not differentiate among excitation mechanisms of the gas,
such as heating by shocks, far-ultraviolet (FUV), or X-rays and cosmic rays,
hence do not provide information about the underlying physics.

The next level of complexity involves modeling the gas as equilibrium photo-dissociation
regions, PDRs \citep{tielenshb1985,Hollenbach1999,rolling07}. These have been
successfully applied to star forming regions and
star-bursts.  However, modeling of Herschel and other observations for, e.g., NGC 253,
NGC 6240 and M82, using these PDRs show that other heating source rather than FUV are
required to reproduce observational data. In
particular, such heating source can be identified in AGN or enhanced cosmic ray ionization
\citep[][among many others]{maloney96, komossa03, martin06-1, papadopoulos10,
meijerink13-1, Rosenberg14}, or mechanical heating due to turbulence
\citep{loenen2008, pan2009-1, aalto13}. The latter is usually not included in ordinary PDR models and is
the focus of this paper.

Various attempts have been made in this direction in modeling star-forming
galaxies and understanding the properties of the molecular gas. However, because
of the complexity and resolution requirements of including the full chemistry in
the models, self-consistent galaxy-scale simulations have been limited mainly to
CO \citep{Kravtsov2002, wada05, cubick2008-1, Narayanan2008-1, inti2009-1, xu2010, perez11,
narayanan2011b, shetty2011, Feldmann2012, Narayanan2013-1, olsen16-1},  but see also
e.g. \citep{olsen15-1} for an effort to model $[\rm{C}II]$.

The rotational transitions of CO up to $J=4 -3$ predominantly probe the
properties of gas with densities in the range of 10$^2$ - 10$^5$~\cmt, and with
temperatures from $\sim 10$K to $\sim 50$K. Higher $J$ transitions probe denser and warmer
molecular gas around $\sim 200$K for the $J = 10-9$ transition.  In addition to
high-$J$ CO transitions, low-$J$ transitions of high density tracers such as CS, CN, HCN, HNC
and HCO$^+$, are good probes of cold gas with $n \sim 10^6$~\cmt. 
Having a broad picture on the line emission of these species provides
a full description of the thermal and dynamical state of the dense gas (where strong
cooling and self-gravity dominate).  Thus, potentially unique signatures of
turbulent and cosmic ray/X-ray heating may lie in the line emission of these
species in star-forming galaxies.

In \cite{mvk12, mvk13-a} we studied the effect of mechanical feedback on
diagnostic line ratios of CO, \thco~ and some high density tracers for grids of
mechanically heated PDR models in a wide parameter space relevant to quiescent
disks as well as turbulent galaxy centers.  We found that molecular line ratios
for CO lines with $J \le 4 -3$ are good diagnostics of mechanical
heating.  In this paper, we build on our findings in \cite{mvk13-a} to apply
the chemistry models to the output of simulation models of star forming
 galaxies, using realistic assumptions
 on the structure of the ISM on unresolved, sub-grid, scales.
mode to construct CO and \thco~ maps for transitions up to $J =
4 - 3$.  Our approach is similar to that by \cite{perez11} where the sub-grid
modeling is done using PDR modeling that includes a full chemical network based
on \cite{umist1999}, which is not the case for the other references mentioned
above.  The main difference of our work from \cite{perez11} is that our sub-grid
PDR modeling takes into account the mechanical feedback in the heating budget;
on the other hand we do not consider X-ray heating effects due to AGN.
The synthetic maps are processed in a fashion that simulates what observers
would measure. These maps are used as a guide to determine how well diagnostics
such as the line ratios of CO and \thco~ can be used to constrain the presence
and magnitude of mechanical heating in actual galaxies.

In the method section we start by describing the galaxy models used, although
our method is generally applicable to other grid and SPH based simulations.  We
then proceed by explaining the procedure through which the synthetic molecular
line emission maps were constructed.  In the results section we study the
relationship and the correlation between the luminosities of CO, \thco, and
H$_2$.  We also present maps of the line ratios of these two molecules and see
how mechanical feedback affects them, how well the physical parameters of the
molecular gas can be determined, when the gas is modeled as a single PDR with
and without mechanical feedback.  In particular, we try to constrain the local
average mechanical heating rate, column density and radiation field and compare
that to the input model.  We finalize with a discussion and conclusions.

\section{Methods \label{sec:paper3_methods}}

In order to construct synthetics emission maps of galaxies, we need two
ingredients: (1) A model galaxy, which provides us with the state of the gas,
and (2) a prescription to compute the various emission of the species.  We start
by describing the galaxy models in Section-\ref{subsec:paper3_galaxymodels}
along with the assumptions used in modeling the gas. The parameters of the gas,
which are necessary to compute the emission of the species and the properties of
the model galaxies chosen, are described in Section-\ref{subsubsec:ingredients}.
  In Section-\ref{subsec:paper3_emission}, we describe the method with which the
sub-grid PDR modeling was achieved and from which the emission of the species
were consequently computed. Sub-grid modeling is necessary since simulations
which would resolve scales where H/H$_2$ transitions occur
\citep{tielenshb1985}, and where CO forms, need to have a resolution less than
$\sim 0.01$ pc.  This is not the case for our model galaxies, but this is
achieved in our PDR models.  The procedure for constructing the emission maps is
described in Section-\ref{subsec:paper3_datacube}.

\subsection{Galaxy models} \label{subsec:paper3_galaxymodels}
In this paper, we will use the data of model galaxies of \cite{inti2009-1}, which are
TreeSPH simulations of isolated dwarf galaxies containing gas stars and dark matter
in a (quasi-) steady state.  The
simulation code calculates self-gravity using a Oct-tree method \citep{barnes86}
and gas dynamics using the softened particle hydrodynamics (SPH) formalism
\citep[see e.g.][]{monaghan92}, in the conservative formulation of
\cite{springelHernquist02}.  It uses an advanced model  for the interstellar medium (ISM), a star 
formation recipe based on a Jeans mass criterion, and a well-defined feedback
prescription.  More details of the code and the simulations can be found in
\cite{inti2009-1}. Below we will give the main ingredients.

The ISM model used in the dynamic simulation is similar, albeit simplified,
to that of \cite{wolfire1995-1, wolfire03}.
It solves for the thermal evolution of the gas including a  range of collisional
cooling processes, cosmic ray heating and ionization. It tracks the development
of the warm neutral medium (WNM) and the cold neutral medium (CNM) HI phases.
The latter is where densities, $n > 10$~\cmt, and low temperatures, $T < 100$ K,
allow the $\rm H_2$  molecules to form.  In violent star-forming galaxies the
time-scale of the variations of the cloud boundary conditions, such as the FUV
irradiation or the external pressure, are fast enough to be comparable to the
time-scale of the HI-$\rm H_2$ phase transition. Hence this transition is handled
in a time-dependent manner by \cite{inti2009-1}.

The FUV luminosities of the stellar particles, which are needed to calculate the
photoelectric heating from the local FUV field, are derived from synthesis
models for a Salpeter IMF with cut-offs at 0.1~\Msun~and 100~\Msun~by
\cite{bruzualCharlot93}, and updated by \cite{bruzualCharlot03}. 
Dust extinction of UV light is not accounted for, other than that from the natal
cloud. For a young stellar  cluster we decrease the amount  of UV extinction
from 75\% to 0\% in 4 Myr \citep [see][]{parravano2003}.

For an estimate of the mechanical heating rate we extract the local dissipative
terms of the SPH equations, the artificial viscosity terms \citep{springel05-1}.
These terms describe the thermalization of shocks and random motions in the gas,
and are in our model ultimately derived from the supernova and the wind energy
injected by the stellar particles that are formed in the simulation
\citep{intiPhdT}.  We realize that this method of computing the local mechanical
heating rate is very approximate.  To be specific: this only crudely models the
actual transport of turbulent energy from large scales to small scales happening
in real galaxies, but for our purposes it suffices to obtain an order of
magnitude estimate of the available energy and its relation with the local star
formation.

We selected two galaxy types from the set of galaxy models by
\cite{inti2009-1}, and applied our PDR models to them.  These galaxies are
star-forming galaxies and have metallicities representing typical dwarfs and
disk-like galaxies, which enable us to study typical star-bursting regions.  The
first model galaxy is a dwarf galaxy with low metallicity ($Z = 0.2$~\zsun). 
The second model is a heavier, disk like, galaxy with metallicity ($Z =
$~\zsun).  The basic properties of the two simulations are summarized in
Table-\ref{tbl:paper3_galaxies}.

\begin{table}[h]
\centering
\begin{tabular}{c c c c c}
  \hline
  abrv & name & mass (\Msun) & Z (\zsun) & gas fraction \\
  \hline
  \hline
  dwarf & coset2 & $10^9$   & 0.2 & 0.5  \\
  disk  & coset9 & $10^{10}$ & 1.0 & 0.2 \\
  \hline
\end{tabular}\caption{Properties of the galaxies used. The gas fraction is the ratio of the gas mass
  relative to the total baryonic mass in the disk.
  (see \cite{inti2009-1} for more detail on modelling each component).
\label{tbl:paper3_galaxies}}
\end{table}

\subsection{Ingredients for further sub-grid modeling}
\label{subsubsec:ingredients}

While our method of constructing molecular emission maps is generally applicable
to grid based or SPH hydrodynamic simulations of galaxies, we use the
simulations of \cite{inti2009-1} since they provide all the necessary
ingredients for our sub-grid modeling prescription.
These are necessary for the post-processing of the snapshots of the hydrodynamic
density field, and to produce realistic molecular line emission maps.  Our
method is applicable to any simulation if it provides a number of physical
quantities for each of the resolution elements (particles, grid cells): the
densities resolved in the simulation must reach $n \sim 100-1000$~\cmt in-order
to produce reliable CO and \thco~ emission maps up to $J = 4 -3$, and the
simulation must provide estimates of the gas temperature, far-UV field
flux and local mechanical heating rate. In essence, the simulation must provide
realistic estimates of the CNM environment in which the molecular clouds
develop. 
In the next section, we describe in detail the assumptions with which
the CNM was modeled. A number of such galaxy models exist (see the introduction
for references),
and with the increase in computing power, more simulations, also in a 
cosmological context, are expected to become available. \cite{inti2009-1}
present a suite of SPH models of disk and dwarf galaxies.
We note that the gas temperature estimated from the PDR models is not the same
as the gas temperature of the SPH particle from the simulation.  The reason for
this is the assumption that the PDR is present in the sub-structures of the SPH
particle.  This sub-structure is not resolved by the large scale galaxy
simulations, hence its thermal state is not probed.  The thermal state of the ISM
depends also on its structure that is known to be clumpy and with a fractal profile
\citep[e.g.][]{hopkins12-1, hopkins12-2, hopkins13-1}.
In star-bursting galaxies the gas density has a continuous distribution and is 
thought to be super-sonically turbulent \citep{norman96-1, Goldman12-1}. Part of the turbulent energy is absorbed
back into the ISM, thus affecting its thermal balance. However the fraction of absorbed
turbulent energy into the ISM is under debate, where a commonly used fraction is about
10\%\citep[e.g.][]{loenen2008}. For more details on sources of mechanical heating and turbulence, and gas
dynamics see \cite{mvk12, mvk13-a} and \cite{inti2009-1} and references therein.
We compared the PDR surface
temperatures with those of the SPH particles as a check for the SPH-determined
temperatures giving good boundary conditions to the embedded PDRs and found good
agreement between them.  We want to stress that we can apply our methodology to
any simulation that fulfills the above criteria.

\subsection{Sub-grid PDR modeling in post-processing mode}
\label{subsec:paper3_emission}

The two main assumptions for the sub-grid modelling are: (1) local dynamical
and chemical equilibrium and (2) that the substructure where H$_2$ forms
complies to the scaling relation of \cite{larson1981}, from which the prescription, by
\cite{inti2006-1}, of the mean $A_V$ given in Eq-\ref{eq:paper3_AV} is derived.
\begin{equation} \label{eq:paper3_AV}
  <A_V> = 3.344 Z \left( \frac{P_e/k_B}{10^4~{\rm cm}^{-3} K} \right)^{1/2}
\end{equation}
$Z$ is the metallicity of the galaxy in terms of \zsun, $P_e$ is the boundary
pressure of the SPH particle and $k_B$ is the Boltzmann constant.  Using the
boundary conditions as probed by the SPH particles and this expression for the
mean $A_V$, we proceed to solve for the chemical and thermal
equilibrium using PDR models.

We assume a 1D semi-infinite plane-parallel geometry for the PDR models whose
equilibrium is solved for using the Leiden PDR-XDR code \cite{meijerink2005-1}. 
Each semi-infinite slab is effectively a finite slab illuminated from one side
by an FUV source.  This is of course an approximation, where the contribution of
the FUV sources from the other end of the slab is ignored, and the exact
geometry of the cloud is not taken into account.
The chemical abundances of the species and the thermal balance along the slab
are computed self-consistently at equilibrium, where the UMIST chemical
network \citep{umist1999} is used.  In this paper we keep the elemental
abundance ratio of $^{12}{\rm C}/^{13}{\rm C}$ fixed to a value of 40 \citep{wilson1994-1},
that is the lower limit of the suspected range in the Milky way.  This ratio is
important in the optically thin limit of the CO line emission at the edges of
the galaxies.  It plays a less significant role in the denser central regions of
galaxies. The same cosmic ray ionization rate used in modelling galaxies was used
in the PDR models.
We have outlined the major assumptions of the PDR models used in this paper, for
more details on these see \cite{meijerink2005-1} and \cite{mvk12, mvk13-a}.

The main parameters which determine the intensity of the emission of a PDR 
are the gas number density ($n$), the FUV flux ($G$) and the depth of the cloud
measured in $A_V$.  In our PDR models we also account for the mechanical
feedback that is introduced as a uniform additional source of heating to the 
thermal budget of the PDR \citep{mvk12}. We refer to these models as mechanically heated 
PDRs (mPDR),  where the additional mechanical heating affects the chemical abundances
of species \citep{loenen2008, mvk12}, as well as their emission \citep{mvk13-a}.  Hence
the fourth important parameter required for our PDR modeling is the mechanical
feedback (\gm).  In addition to these four
parameters, metallicity plays an important role, but this is taken constant
throughout each model galaxy (see Table-\ref{tbl:paper3_galaxies}).  The SPH
simulations provide local values for $n$, $G$, and \gm.

Based on the PDR model grids in \cite{mvk13-a}, we can compute the line emission
intensity of CO and \thco~ given the four parameters ($n$, $G$, \gm, $A_V$) at a
given metallicity.
We briefly summarize the method by which the emission was computed by
\cite{mvk13-a}.  For each PDR model, the column densities of CO and \thco, the
mean density of their main collision partners, H$_2$, H, He and e$^-$, and the
mean gas temperature in the molecular zone (i.e. the region in the cloud where species
are prevalently present in molecular form, in general beyond $A_{\rm V} = 5~~{\rm mag}$)
are extracted from the model grids.
Assuming the LVG approximation, these quantities are used as input to RADEX
\citep{radex} which computes the line intensities.
In the LVG computations a micro-turbulence line-width, $v_{\rm turb}$, of 1 \kms
was used.
Comparing this line width to the velocity dispersion of the SPH particles, we
see that the velocity dispersion for particles with $n > 10$~\cmt, where most of
the emission comes from and which have $A_V > 5$~mag, is $\sim$1 \kms.
The choice of the micro-turbulence line-width does not affect the general
conclusions of the paper. This is discussed in more detail in
Section-\ref{subsec:improvementsandprospects}.

In this paper, the parameter space used by \cite{mvk13-a} is extended to
include $10^{-3} < n< 10^{6}$~\cmt, $10^{-3} < G < 10^{6}$, where $G$ is
measured in units of $G_0 = 1.6 \times 10^{-3}$~\inten. Moreover the emission
is tabulated for $A_V = 0.01, 0.1, 1 ... 30$~mag.  The range in \gm is wide
enough to cover all the states of the SPH particles.  For each emission line of
CO and \thco~ we construct 4D linear interpolation tables from the $\log_{10}$ of $n$, $G$ and \gm. 
The dimensions of these tables is ($\log_{10}(n), \log_{10}(G), \log_{10}(\Gamma_{\rm mech}), A_{\rm V}$) $ = (20, 20, 24, 22)$.  
Consequently, given any
set of the four PDR parameters for each SPH particle, we can compute the
intensity of all the lines of these species.  About 0.1\% of the SPH particles
had their parameters outside the lower bounds of the interpolation tables,
mainly $n < 10^{-3}$\cmt~and $G < 10^{-3}$.  The disk galaxy consists of $2
\times 10^6$ particles, half of which contribute to the emission. The surface
temperature of the other half is larger than $10^4$K, which is caused by high
\gm where no transition from H to H$_2$ occurs in the PDR, thus CO and \thco~
are under-abundant.  We ignore these SPH particles since they do not contribute
to the mean flux of the emission maps and the total luminosities.

The use of interpolation tables in computing the emission is because of CPU time limitations.
Computing the equilibrium state for a PDR model consumes, on average, 30
seconds on a single core$^{[}$\footnote{The PDR code is executed on an Intel(R) Xeon(R) W3520 processor
and compiled with gcc 4.8 using the -O3 optimization flag.}$^{]}$.  Most of the time, about 50\%, is spent in computing the equilibrium
state up to $A_V = 1$~mag near the H/H$_2$ transition zone.  Beyond $A_V =
1$~mag, the solution advances much faster.  Finding the equilibria for a large
number of SPH particles requires a prohibitively long time, thus we resort to
interpolating.  Although interpolation is less accurate, it does the required
job.  On average it takes 20 seconds to process all the SPH gas particles with $n > 10^{-3}$~\cmt
and produce an emission map for each of the line emission of CO and \thco, with the
scripts running serially on a single core.

\subsection{Construction of synthetic emission maps and data cubes}
\label{subsec:paper3_datacube}

The construction of the flux maps is achieved by the following steps:
\begin{enumerate}
\item Construct a 2D histogram (mesh) over the spatial region of interest.
\item For each bin (grid cell, pixel) compute the mean flux in units of energy per unit
 time per unit area.
\item Repeat steps 1 and 2 for each emission line.
\end{enumerate}

In our analysis the region of interest is $R < 8$~kpc for the disk galaxy and $R < 2$~kpc for
the dwarf galaxy.  As for the pixel size, choosing a mesh with $100 \times 100$ grid cells, results in
having on the order of 100 SPH particles per grid cell that is a statistically significant distribution
per pixel.  The produced emission maps with such a resolution have smooth profiles for our galaxies 
(See Section-\ref{subsec:paper3_emission-map-construction} for more details).
In practice a flux map is constructed from the brightness temperature of a line,
measured in K, that is spectrally resolved over a certain velocity range.  This
provides a spectrum at a certain pixel as a function of velocity.  The
integrated spectrum over the velocity results in the flux.  The velocity
coordinate, in addition to the spatial dimensions projected on the sky, at every
pixel can be thought of as a third dimension; hence the term ``cube''.  In what follows 
we describe the procedure by which we construct the data cube 
for a certain emission line from the SPH simulation.  Each SPH particle has a different
line-of-sight velocity and a common FWHM micro-turbulence line width of 1 \kms. 
By adding the contribution of the Gaussian spectra of all the SPH particles
within a pixel, we can construct a spectrum for that pixel.  This procedure can
be applied to all the pixels of our synthetic map producing a synthetic data
cube.  The main assumption in this procedure is that the SPH particles are
distributed sparsely throughout each pixel and in the line of sight velocity space.
We can estimate the number density of SPH particles per pixel per line-of-sight
velocity bin by considering a typical pixel size of $\sim 1$ kpc\footnote{A pixel size
of 1'' on the sky corresponds to a 1 kpc object that is $\sim 3.6$~Mpc away. Such an object
can be easily resolved by e.g. ALMA where a resolution of $\sim 0.1$'' is now routinely reached
at almost all the frequency bands it operates in.}
and a velocity
bin equivalent to the adopted velocity dispersion.  The typical range in line of
sight velocities in both simulations ranges from -50 to +50 km/s, which results 
in 100 velocity bins. For reference, the line-of-sight velocity dispersion in
star-bursting galaxies could be as high as 500 km/s. But our model galaxies 
are smaller and less violent, resulting in lower line-of-sight velocities.
With an average number of 5000 SPH particles in a pixel, the number density of
SPH particles per pixel per velocity bin is 50.  The scale size of an SPH
particle is on the order of $\sim 1$ pc, which is consistent with the size
derived from the scaling relation of Eq-1 by \citep{larson1981} by using a
velocity dispersion of 1 km/s.  Thus, the ratio of the projected aggregated area
of the SPH particles to the area of the pixel is $\sim 10^{-4}$.  This can
roughly be thought of as the probability of two SPH particles overlapping along
the line of sight within 1 km/s.

\section{Results}

\subsection{Emission maps} \label{subsec:paper3_emission-map-construction}
\begin{figure*}[!tbh]
   \center
   \includegraphics[scale=1.0, bb=54 284 558 507]{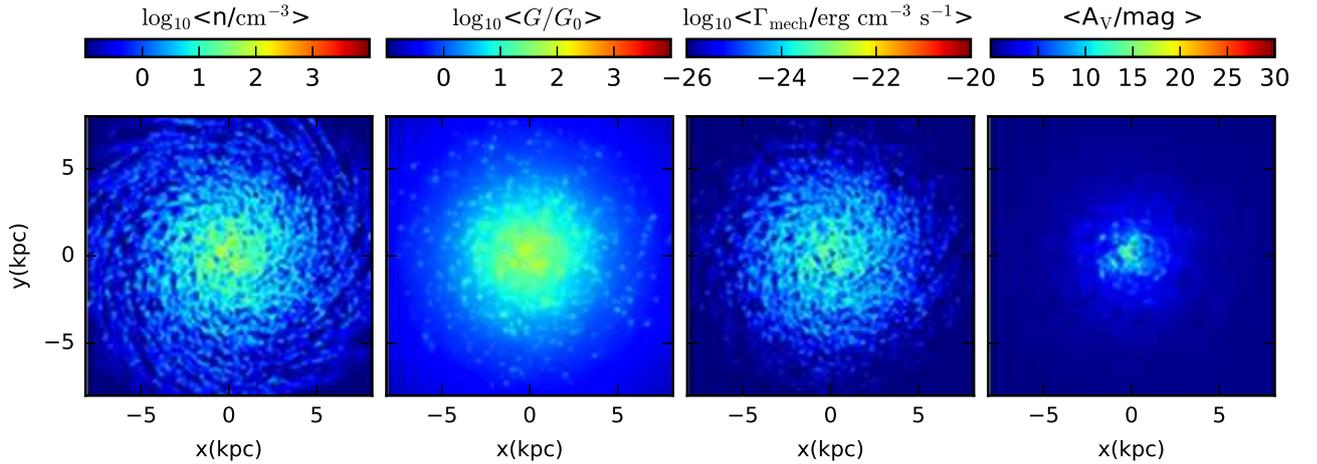}
   \caption{{\bf Left to right:} distribution maps of the gas
     density, FUV flux, mechanical heating rate and the $A_V$ of the model
     disk galaxy. The galaxy is viewed face on where
     the averages (except $A_{\rm V}$) are computed by averaging along
     the line of sight.
    \label{fig:paper3_n_g_gmech_av_maps_disk}}
\end{figure*}

\begin{figure*}[!tbh]
   \center
   \includegraphics[scale=1.0, bb=28 25 376 315]{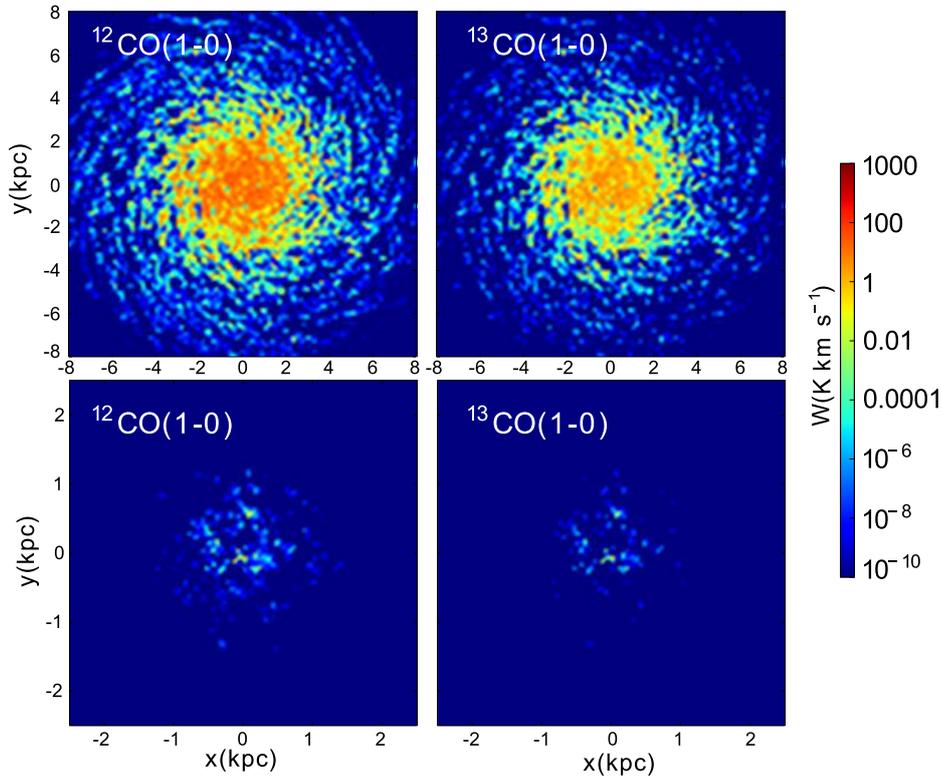}
   \caption{ (Top row) CO($1-0$) and \thco(1-0) fluxes of the disk galaxy.
    (Bottom row) CO(1-0) and \thco(1-0) emission for the dwarf galaxy.  The 
    pixel size in these maps is 0.16 $\times$ 0.16 kpc.
    \label{fig:paper3_sample-emissions-maps}} 
\end{figure*}

In Figure-\ref{fig:paper3_n_g_gmech_av_maps_disk}, we show the distribution map 
of the disk galaxy for the input quantities to PDR models 
from which the emission maps are computed. The analogous maps for the dwarf galaxy
are show in Figure-\ref{fig:paper3_n_g_gmech_av_maps_dwarf} in the appendix.
The emission maps of the first rotational transition of CO and \thco, for the model 
disk and dwarf galaxy are shown in Figure-\ref{fig:paper3_sample-emissions-maps}.  These 
maps were constructed using the procedure described in
Section-\ref{subsec:paper3_emission}.  As the density and temperature of the gas increase towards the
inner regions, the emission is enhanced.
Comparing the corresponding top and bottom panels of
Figure-\ref{fig:paper3_sample-emissions-maps}, we see that the emission of the
dwarf galaxy is significantly weaker than that of the disk galaxy.  The gas mass
of the dwarf galaxy is four times less than the disk galaxy's. Moreover, the
metallicity of the dwarf galaxy is 5 times lower than that of the disk galaxy. 
Hence the column densities of CO and \thco~ are about 100 times lower in the
former galaxy.
The mean gas temperatures used to compute the emission is $\sim 10$ times lower
in the PDR sub-grid modeling of the dwarf galaxy compared to the disk galaxy,
which results in weak excitation of the upper levels of the molecules through
collisions.  All these factors combined result in a reduced molecular luminosity
in the dwarf galaxy, which is $\sim 10^4$ times weaker than that of the disk
galaxy.

\begin{figure*}[!tbhp]
   \center
    \includegraphics[scale=0.7, bb=0 0 504 180]{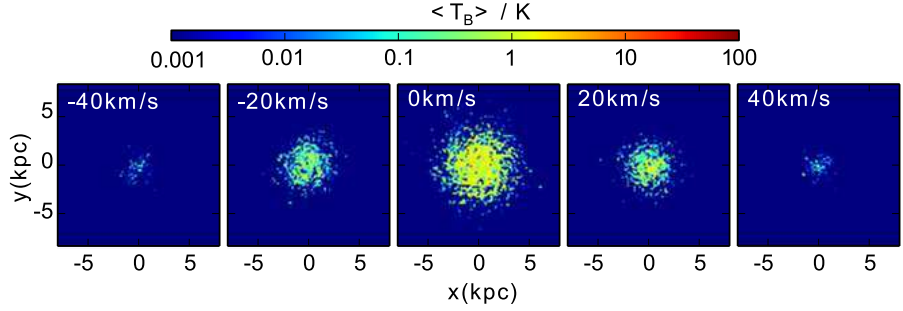}
  \vspace{0.0cm}
  \caption{Channel maps of the CO($1-0$) emission of the disk galaxy.  The width
  of each velocity channel is 20 \kms, where the centroid of the velocity channel, 
  is indicated at the top of each panel. Since the 
  galaxy is projected face-on at the sky, most of the emission 
  emanates from the channel [-10, 10] \kms centered at $V_{\rm los} = 0$ \kms.
  \label{fig:paper3_channel_map}}
\end{figure*}

We demonstrate the construction of the data cube, described in
Section-\ref{subsec:paper3_datacube}, by presenting the CO$(1 - 0)$ emission map
of the disk galaxy in Figure-\ref{fig:paper3_channel_map}.
These maps provide insight on the velocity distribution of the gas along the
line-of-sight.
In the coordinate system we chose, negative velocities correspond to gas moving
away from the observer, where the galaxy is viewed face-on in the sky.  Thus,
the velocities of the clouds are expected to be distributed around a zero mean. 
The width in the velocity distribution varies depending on the spatial location
in the galaxy.  For example,  at the edge of the galaxy the gas is expected to
be quiescent, with a narrow distribution in the line-of-sight velocities
($V_{\rm los}$).
This is seen clearly in the channel maps $|V_{\rm los}| = 20$~\kms, where the
CO$(1 - 0)$ emission is too weak outside the $ R > 3$~kpc region.  In contrast
the emission of these regions are relatively bright in the $|V_{\rm los}| =
0$~\kms map.  In the inner regions, $ R < 1$~kpc, the CO$(1 - 0)$ emission is
present even in the 40 \kms channel, which is a sign of the wide dynamic range
in the velocities of the gas at the central parts of the galaxy.

\begin{figure*}[!tbhp]
   \center
    \includegraphics[scale=0.42, bb=-0 -0 864 288]{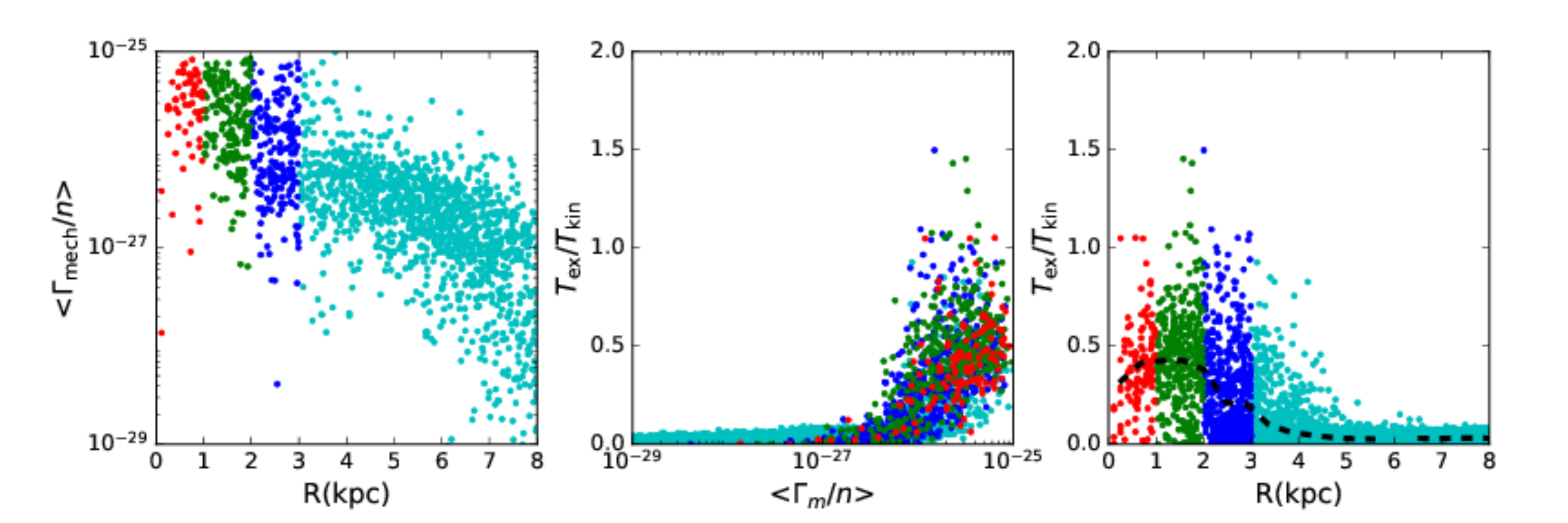}
  \caption{({\bf Left}) Mean mechanical heating per hydrogen (in erg s$^{-1}$) nucleus as a
  function of the
    distance from the center of the disk galaxy.({\bf Center}) The ratio of the
    mean excitation temperature of the CO(1-0) line to mean kinetic temperature
    as a function of mechanical heating rate per hydrogen nucleus. ({\bf Right})
    The ratio of the mean excitation temperature of the CO(1-0) line to mean
    kinetic temperature as a function of the distance from the center of the
    disk galaxy.  The different colors correspond different galactocentric
    distance intervals.
    Red, green blue and cyan correspond to intervals [0, 1], [1,2] [2,3] and [3,
    8] kpc respectively.\label{fig:paper3_Tex}}
\end{figure*}

The relationship between excitation temperature, $T_{\rm
ex}$ of the CO($1-0$) line, the distance of
the molecular gas from the center of the galaxy ($R$) and mechanical feedback 
is illustrated in Figure-\ref{fig:paper3_Tex};  we plot the averages of \gm$/n$, the mechanical
heating rate per H nucleus, and the mean $T_{\rm ex}$ of the SPH particles in
each pixel of the emission maps.  $T_{\rm ex}$ for each SPH particle is a by
product of the RADEX LVG computations \citep{radex}.  We highlight the two
obvious trends in the plots: (a) The mechanical heating per H nucleus increases
as the gas is closer to the center, (b) The excitation temperature correlates
positively with mechanical feedback and correlates negatively with distance from
the center of the galaxy.  It is also interesting to note that, on average, the
SPH particles with the highest \gm$/n$~have the highest excitation temperatures
and are the closest to the center, see red points in the middle panel of
Figure-\ref{fig:paper3_Tex}.  On the other hand, gas situated at $R > 3$~kpc has
average excitation temperatures less than 10 K and approaches 2.73 K, the cosmic
microwave background temperature we chose for the LVG modeling, at the outer
edge of the galaxy.  This decrease in the excitation temperature is not very
surprising, since at the outer region CO is not collisionally excited due to
collisions with H$_2$, which has a mean abundance 10 times lower than that of
the central region.  Collisional excitation depends strongly on the kinetic
temperature of the gas.  In \cite{mvk12}, it was shown that small amounts of
\gm~are required to double the kinetic temperature of the gas in the molecular
zone, where most of the molecular emission originate. However, \gm~is at least
100 times weaker in the outer region compared to the central region, which
renders mechanical feedback ineffective in collisionally exciting CO.

\subsection[The X factor: correlation between CO and H$_2$]{The X factor:
correlation between CO emission and H$_2$ column density}

\begin{figure}[!tbhp]
  \center
    \includegraphics[scale=0.8, bb=0 0 288 288]{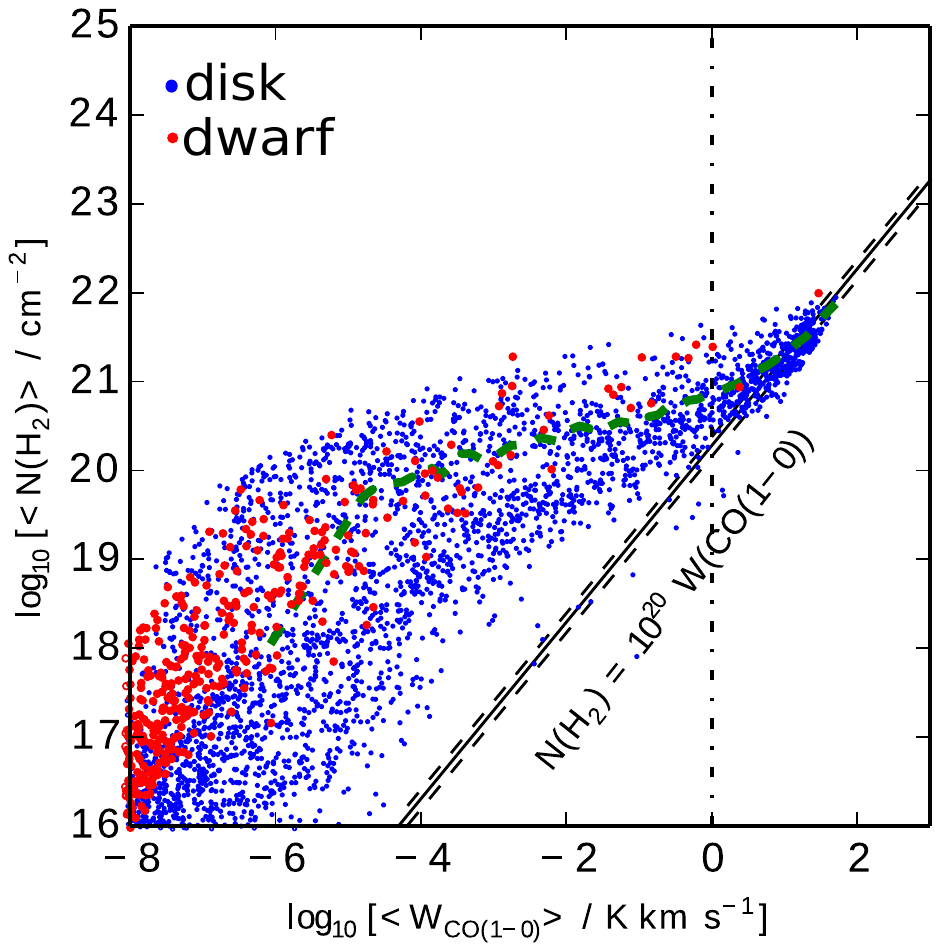}
  \caption{\nhtwo vs. CO(1-0) flux of the synthetic emission maps for the disk and dwarf
    galaxies. The blue and red points correspond to the pixels of the disk and dwarf
    galaxies respectively.  The solid black line is the $\wco = \xfcomw$~\nhtwo curve, with
  the observed $\pm 30\%$ uncertainty band \citep{bolatto2013-1} shown by the
  black dashed lines. This uncertainty could be up to a factor of two under a
  variety of conditions.  $\xfco$ for pixels with $\wco < 10$~\Kkms diverges
  from that of the Milky way, where the mean $\xfco$ for the pixels is plotted
  in green.  We note that 99\% of the luminosity of the disk galaxy emanates
  from pixels whose $\wco > 1$~\Kkms indicated by the dot-dashed line.
  \label{fig:paper3_xfactor}} 
\end{figure}

Since H$_2$ can not been observed through its various transitions in cold
molecular gas whose $T_{{\rm kin}} < 100$ K,  astronomers have been using
the molecule CO as a proxy to derive the molecular mass in the ISM of galaxies. It has
been argued that the relationship between CO(1-0) flux and $N(\rm H_2)$ is more
or less linear \citep[][and references therein]{solomon1987-1, bolatto2013-1}
with $X_{\rm CO} = N({\rm H}_2) / \wco$ nearly constant, where the
proportionality factor $\xfco$ is usually referred to as the $X$-factor.
The observationally determined Milky Way $\xfco$, $\xfcomw$, is given by $\sim 2
\times 10^{20}$~cm$^{-2}$~(\Kkms)$^{-1}$ \citep[][]{solomon1987-1}.
By using the emission map of CO(1-0) presented in
Figure-\ref{fig:paper3_sample-emissions-maps}, and estimating the mean \nhtwo
throughout the map from the PDR models, we test this relationship in
Figure-\ref{fig:paper3_xfactor}.

It is clear that only for pixels with $\wco > 10$~ \Kkms $\xfco$ approaches
$\xfcomw$.
These pixels are located within $R \lesssim 2$~kpc of the disk galaxy, and $R
\lesssim 0.2$~kpc of the dwarf galaxy.  Whenever $\wco < 10$~ \Kkms, $\xfco$
increases rapidly reaching $\sim 1000 \xfcomw$.  Looking closely at pixels
within $\wco$ intervals of $[0.1, 1]$, $[1, 10]$ and $> 10$ \Kkms we see that
the gas average densities in these pixels are $\sim 20, 80$ and 300~\cmt
respectively.  This indicates that as the gas density becomes closer to the
critical density, $n_{\rm crit}$\footnote{We use the definition $n_{\rm crit}
\equiv A_{ij}  / K_{ij}$~\citep[cf.][]{tielens2005physics}, where $K_{ij}$ is
the collisional rate coefficient of the transition from the $i^{th}$ to the
$j^{th}$ level and $A_{ij}$ is the spontaneous de-excitation rate, the Einstein
A coefficient. See Krumholz 2007 for the modified definition of the
critical density which takes self-shielding into account}, of the
CO(1-0) transition, which is $\sim 2 \times 10^3$\cmt,
$\xfco$ converges to that of the Milky Way.  This is not surprising since as the
density of gas increases, the mean $A_V$ of an SPH particle increases to more
than 1 mag.  In most cases, beyond $A_V > 1$~mag, most of the H and C atoms are
locked in H$_2$ and CO molecules respectively, where their abundances become
constant.  This leads to a steady dependence of the CO emission on $A_V$, and
hence H$_2$.  This is not the case for $A_V < 1$~mag, where strong variations in
the abundances of  H$_2$ and CO result in strong variations in the column
density of H$_2$ and the CO emission, leading to the spread in $\xfco$ observed
in Figure-\ref{fig:paper3_xfactor}.  A more precise description on this matter
is presented by \cite{bolatto2013-1}.

Most of the gas of the dwarf galaxy lies in the $A_V < 1$~mag range.  Moreover,
the low metallicity of dwarf galaxy results in a smaller abundance of CO in
comparison to the disk galaxy, and thus a lower $\wco$.  This results in an
$\xfco$ which is 10 to 100 times higher than that of the Milky Way
\citep{leroy2011-1}.
Our purpose of showing Figure-\ref{fig:paper3_xfactor} is to check the validity
of our modeling of the emission.  \cite{maloney98} provide a more rigorous
explanation on the $\wco$ and \nhtwo relationship.

\subsection{Higher $J$-transitions} \label{subsec:paper3_higher-J}

\begin{figure}[!tbh]
   \center
    \includegraphics[scale=0.8, bb=162 252 450 540]{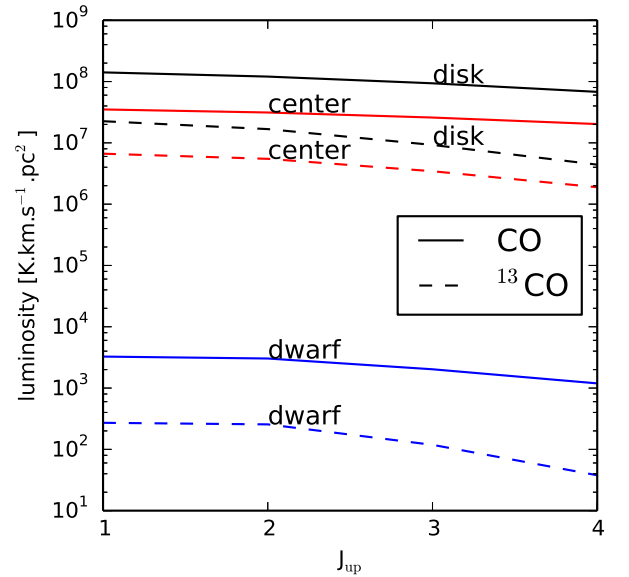}
  \caption{Total luminosity of CO and \thco~transitions up to $J=4-3$ of the
  disk galaxy (black curves), the $R < 2$~kpc region of the disk galaxy
  (red curves) and the dwarf galaxy (blue curves). 
    \label{fig:paper3_total_luminosity}}
\end{figure}

So far we have only mentioned the CO(1-0) transition.  The critical density of
this transition is $2 \times 10^3$~\cmt.  The maximum density of the gas in our
simulations is $\sim 10^4$~\cmt.  It is necessary to consider higher $J$
transitions in probing this denser gas.  The critical density of the CO(4-3)
transition is $\sim 10^5$~\cmt, which corresponds to densities 10 times higher
than the maximum of our model galaxies.
Despite this difference in densities the emission of this transition and the
intermediate ones, $J = 2-1$ and $J = 3-2$, are bright enough to be observed due
to collisional excitation mainly with H$_2$ .  To have bright emission from
these higher $J$ transitions, it also necessary for the gas to be warm enough,
with $T_{\rm kin} \gtrsim 50$~K, so that these levels are collisionally
populated.  In Figure-\ref{fig:paper3_total_luminosity}, we show the
luminosities of the line emission up to $J = 4-3$ of CO and \thco, emanating
from the disk and dwarf galaxies.
In addition to that, we present the luminosity of the inner, $R < 0.5$ kpc,
region of the disk galaxy where \gm~is enhanced compared to the outer parts.
This allows us to understand the trend in the line ratios and how they spatially
vary and how mechanical feedback affects them (see next section).  The ladders
of both species are somewhat less steep for the central region, which can be
seen when comparing the black and red curves of the disk galaxy in
Figure-\ref{fig:paper3_total_luminosity}.  Hence, the line ratios for the
high-$J$ transitions to the low-$J$ transitions are larger in the central
region.  This effect of \gm~was also discussed by Kazandjian {\it et al} (2015),
who also showed that line ratios of high-$J$ to low-$J$ transitions are enhanced
in regions where mechanical heating is high, which is also the case for the
central parts of our disk galaxy.
We will look at line ratios and how they vary spatially in
Section-\ref{sec:paper3_diagnostics}.

\subsection{Diagnostics} \label{sec:paper3_diagnostics}

In this section we use synthetic emission maps, for the $J > 1-0$ transitions,
constructed in a similar fashion as described in
Section-\ref{subsec:paper3_emission-map-construction}.
These maps are used to compute line ratios among CO and \thco~ lines where we
try to understand the possible ranges in diagnostic quantities.  This can help
us recover physical properties of a partially resolved galaxy.

\begin{figure*}[!tbhp]
   \center
    \includegraphics[scale=0.5, bb=0 0 718 742]{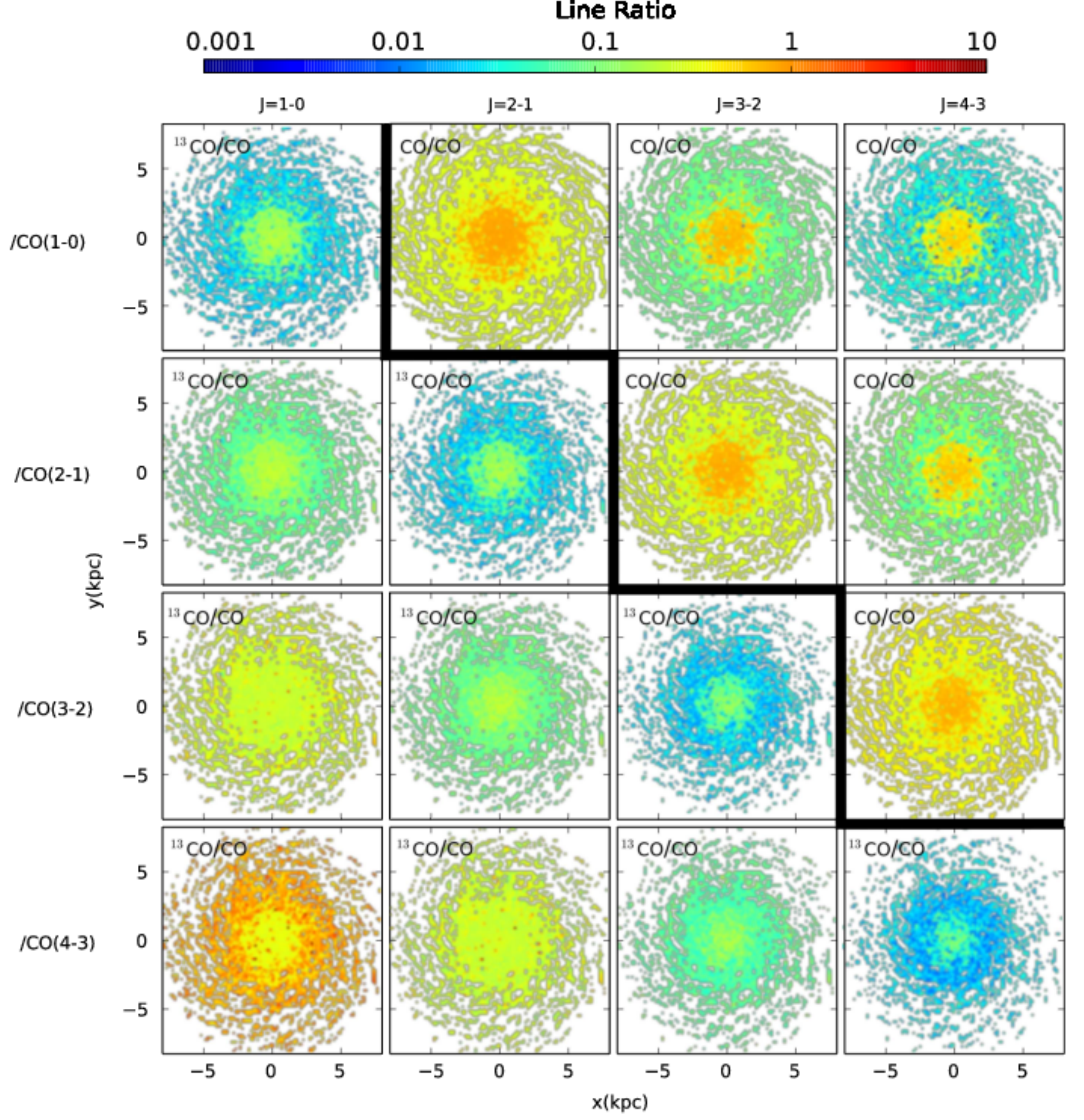}
  \caption{Line ratio maps for various transitions of CO and \thco~ for the disk
  galaxy. The transition
    of the line in the numerator is specified at the top of each column, whereas
    that of the denominator is specified at the left of each row.  For example,
    the panel in the third row of the second column corresponds to the line
    ratio map of \thco(2-1)/CO(3-2); the species involved in the line ratio are
    specified at the top left corner of each panel.  Ratios larger than unity
    are typical to the central regions $R < 2$~kpc.  Line ratio maps between CO
    transitions are to the right of the zig-zagged line, whereas the remaining maps are
    for line ratios between \thco~ and CO.  Ratios involving $J \ge 3 - 2$
    transitions trace the small scale structure of the molecular gas for $R >
    2$~kpc. \label{fig:paper3_line_ratio_maps}}
\end{figure*}

In Figure-\ref{fig:paper3_line_ratio_maps}, we show ratio maps for the various
transitions of CO and \thco. In these maps, line ratios generally exhibit
uniform distributions in the regions where $R < 2$ kpc.  This uniform region
shrinks down to 1 kpc as the emission from transition in the denominator becomes
brighter.  This is clearly visible when looking at the corresponding \thco/CO
line rations in the panels along the diagonal of
Figure-\ref{fig:paper3_line_ratio_maps}.  This is also true for the CO/CO
transitions shown in the upper right panels of the same figure.  Exciting the $J
> $ $1 - 0$ transitions requires enhanced temperatures, where \gm~and H$_2$ densities
higher than $10^3$ \cmt plays an important role.  Such
conditions are typical to the central 2 kpc region that result in bright 
emission of $J > 1-0$ transitions and lead to forming the compact peaks within that region
that we mentioned.
The peak value of the line ratios of CO/CO transitions at the center is around
unity, compared to 0.1 - 0.3 for ratios involving transitions of \thco/CO.
Since $J > 1-0$ transitions are weakly excited outside the central region, the
line ratios decrease, e.g., by factors of 3 to 10 towards the outer edge of the
galaxy for CO(2-1)/CO(1-0) and CO(4-3)/CO(1-0), respectively.  Another
consequence of the weak collisional excitation of CO and \thco~ is noticed by
looking again at the \thco/CO transitions along the diagonal panels of
Fig-\ref{fig:paper3_line_ratio_maps}, where the small scale structure of cloud
``clumping'', outside the central region becomes evident by comparing the
\thco(1-0)/CO(1-0) to \thco(4-3)/CO(4-3).  This dense gas is compact and
occupies a much smaller volume and mass, approximately 10\% by mass.

Similar line ratio maps can be constructed for the dwarf galaxy, which are
presented in Figure-\ref{fig:paper3_line_ratio_maps_dwarf} of the Appendix. 
These maps can be used to constrain the important physical parameters of the gas
of both model galaxies, as we will demonstrate in the next section.

\section[Application: modeling extra-galactic sources]{Application: modeling
extra-galactic sources using PDRs and mechanical feedback}
\label{sec:paper3_recovering}

Now that we have established the spatial variation of diagnostic line ratios in
the synthetic maps,  we can use them to recover the physical parameters of the
molecular
 gas that is emitting in CO and \thco.

The synthetic maps that we have constructed assume a high spatial resolution of
100 $\times$ 100 pixels, where the size of our model disk galaxy is $\sim 16$
kpc.  If we assume that the galaxy is in the local universe at a fiducial
distance of 3 Mpc (the same distance chosen by \cite{perez11}), which is also
the same distance of the well known
galaxy NGC 253, then it is necessary to have an angular resolution of 1'' to
obtain such a resolution.  This can be easily achieved with ALMA where it can easily
achieve a resolution of 0.1'' at all the frequency bands it operates in.
A grid of size 21 $\times$ 21 is used to allow for a $\sim 0.8$kpc resolution per pixel,
which is a typical resolution that can be achieved with single-dish studies of 
nearby galaxies such as the HERACLES/IRAM-30m survey \citep{leroy09-1}.

\begin{figure}[!tbh]
   \center
  \includegraphics[scale=0.90, bb=0 1 230 471]{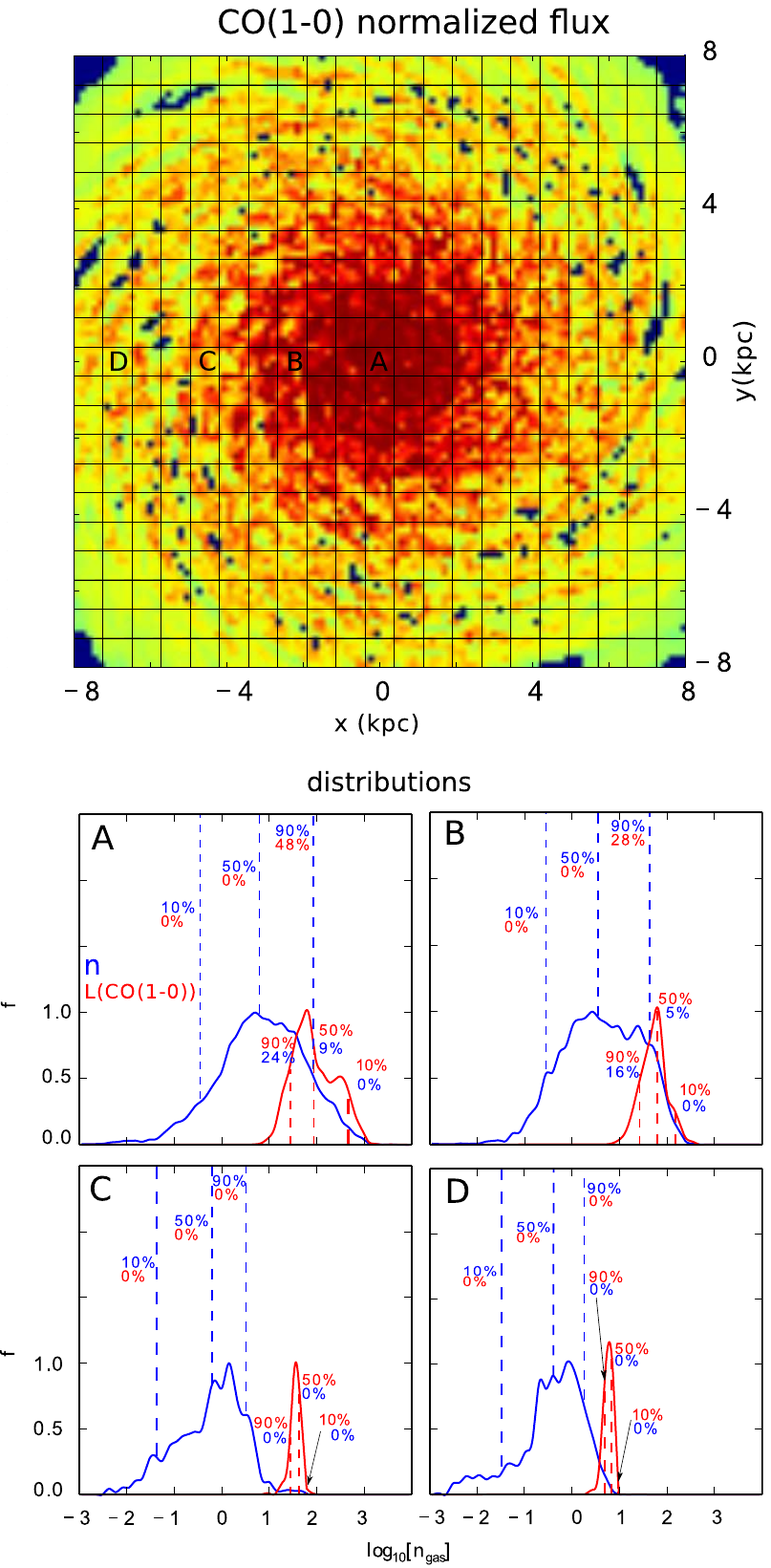}
  \caption{Distributions of gas density and $W({\rm CO}(1-0))$ as a function of
  distance
    from the center of the disk galaxy.  The normalized $\wco$ map, with the
    overlaid grid which has a resolution of $21 \times 21$ pixels is show in the
    {\bf Upper} panel.
    The labels A, B, C, D correspond to the pixels at distances of 0, 2, 4, and
    7 kpc from the center of the galaxy.  In the {\bf bottom} panels we plot, in
    blue, the gas density distribution function, $f(n)$, of the SPH particles in
    these pixels.  The fractional contribution of the SPH particles to the
    CO(1-0) luminosity of the pixel as a function of gas density, $f(L_{{\rm
    CO}(1-0)})(n)$, is plotted in red.
    Both curves are scaled to their maximum value so that we can compare the
    ranges where they overlap.
    The cumulative distribution function is defined as ${\rm CDF}(n) =
    \int_{10^{-3}}^n f(n) d n$.  The blue dashed lines indicate the gas
    densities where the ${\rm CDF} = 10, 50, 90$\% respectively.
    Below these percentages, in red, are the fractional contribution of these
    SPH particles to the luminosity of the pixel.  This contribution is defined
    as ${\rm LCDF}(n) = \int_{10^{-3}}^n f(L_{{\rm CO}(1-0)})(n) d n$.  The red
    dashed lines represent the ${\rm LCDF} = \int_n^{10^{4}} f(L_{{\rm
    CO}(1-0)})(n) d n$ along with the analogous CDF integrated from $n =
    10^4$~\cmt.  The corresponding distribution plots as a function of $G_0$,
    \gm~and $A_V$ are shown in Figure-\ref{fig:paper3_dist-app} in the appendix.
\label{fig:paper3_dist}}
\end{figure}

In the top panel of Figure-\ref{fig:paper3_dist}, we show the normalized $\wco$
map, normalized with respect to the peak flux, with an overlaid mesh of the
resolution mentioned before.  We re-compute the emission maps for all the CO and
the \thco~lines using the 21 $\times$ 21 pixel grid.  Each pixel in the mesh 
contains on average a few thousand SPH particles.  In what follows we treat these
synthetic emission maps as input and try to find the best fitting mPDR models
by following a minimization procedure applied to a certain pixel.  Using the 
emission computed from the PDR models, we estimate the
parameters $n, G, \Gamma_{\rm mech}$ and $A_V$ that best fit the emission of that pixel.
Since we consider transitions up to $J = 4 - 3$ for CO and \thco, we have a total of 8
transitions, hence $8 - 4 = 4$ degrees of freedom in our fits.  The purpose of
favoring one PDR component in the fitting procedure is not to reduce the degrees
of freedom, since for every added PDR component we lose five degrees of
freedom, which could result in fits that are less significant.

\begin{figure}[!tbh]
   \center
  \includegraphics[scale=0.9, bb=0 8 240 466]{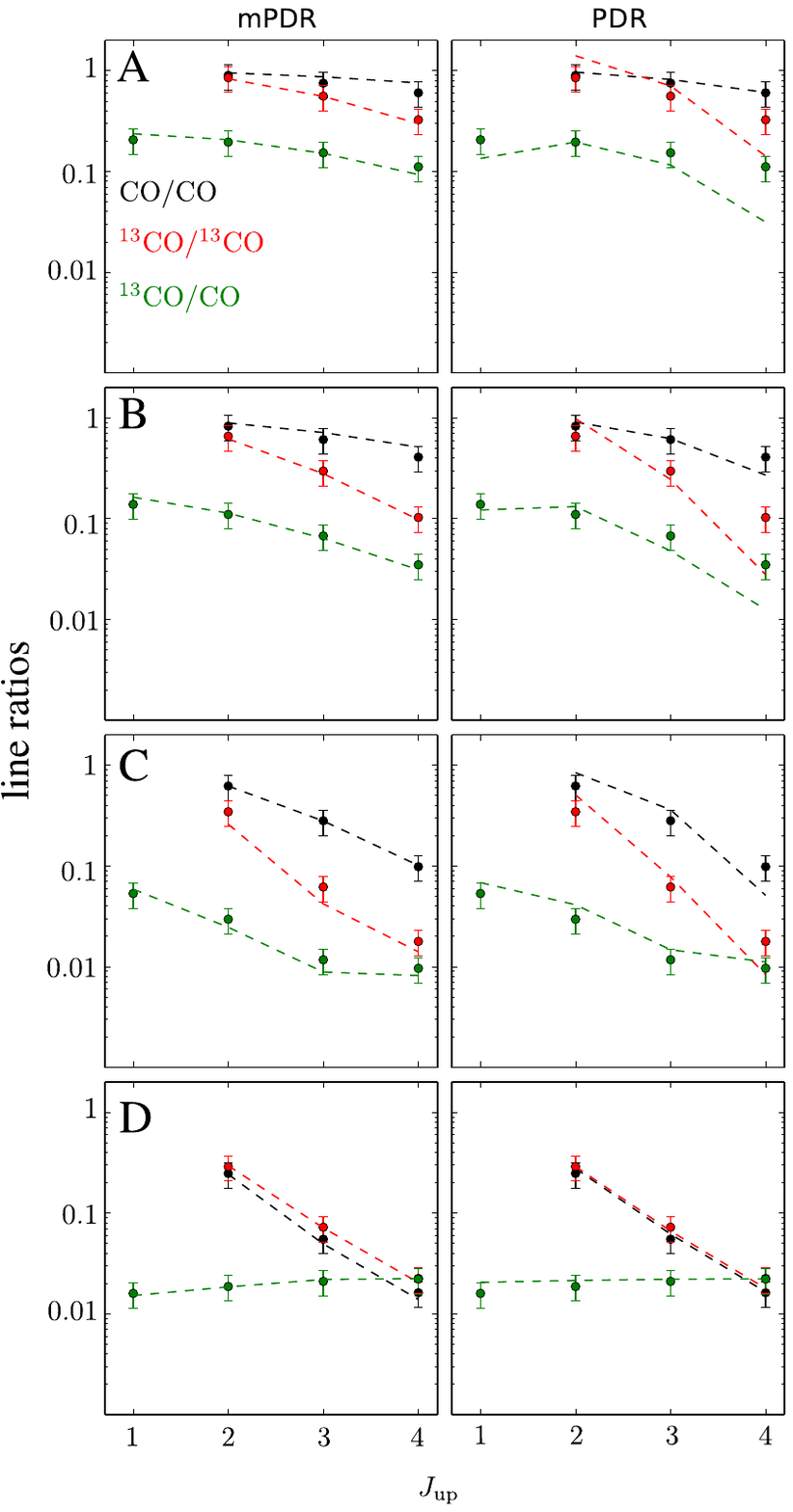}
  \caption{Constraining the parameters for a resolved galaxy.  The black dots
  correspond to the
    line ratios of CO($J \ge 2 - 1$) with CO(1-0) for a pixel in the emission
    map.  The red dots correspond to the line ratios of \thco($J \ge 2 - 1$)
    with \thco(1-0) for the same pixel.  The green dots correspond to the line
    ratios of the same transitions of \thco~ and CO for a pixel.  The dashed
    line are the line ratios of these transitions of the best fit PDR model. 
    The PDR models in the left column labeled ``mPDR''
    consider a range of possible \gm in the heating budget while fitting these
    line ratios (the dots), while the plots labeled ``PDR'' do not take 
    \gm~into account.  The labels A, B, C, and D correspond to the pixels 
    indicated in the top panel of Figure-\ref{fig:paper3_dist}.  The parameters 
    of the best fit PDR models for each pixel are listed in
    Table-\ref{tbl:paper3_comaprison}.
\label{fig:paper3_fitting}}
\end{figure}

The statistic we minimize in the fitting procedure is : 

\begin{equation}
  \chi^2 = \sum_j \sum_i \left[\frac{(r^i_o - r^j_m)}{\sigma^i_o}\right]^2
\end{equation}

\citep{nrcpp08}, where $r_o$ and $\sigma_o$ are the observed values and assumed
error bars of the line ratios of the pixel in the synthetic map.  $r_m$ is the
line ratio for the single PDR model whose parameter set we vary to minimize
$\chi^2$.  The index $i$ corresponds to the different combinations of line
ratios.  The line ratios we try to match are the CO and \thco~ ladders
normalized to their $J = 1 - 0$ transition, in addition to the ratios of \thco~
to the CO ladder. These add up to 10 different line ratios which are not
independent, and the number of degrees of freedom remains 4 for the mechanically
heated model (mPDR), and 3 for the PDR which does not consider mechanical
heating.

In Figure-\ref{fig:paper3_fitting}, we show the fitted line ratios of the pixels
labeled (A, B, C, D) in Figure-\ref{fig:paper3_dist}.  The first row, labeled
{\bf A}, corresponds to the central pixel in the map.  For this pixel, 90\% of
the emission emanates from gas whose density is higher than 10~\cmt,
constituting 24\% of the mass in that pixel.  The normalized cumulative
distribution functions for the gas density (blue curves) and CO(1-0) luminosity
(red curves) are shown in the bottom panels of Figure-\ref{fig:paper3_dist}. 
The blue dashed lines indicates the density where 10, 50 and 90\% of the SPH
particles have a density up to that value.  The numbers in red below these
percentages indicate the contribution of these particles to the total
luminosity of that pixel.  For instance, in pixel {\bf A} 10\% of the particles
have a density less than $\sim 0.03$~\cmt and these particles contribute 0\% to
the luminosity of that pixel;  at the other end,  90\% of the particles have a
density less than $\sim 100$~\cmt and these particle contribute 48\% to the
luminosity of that pixel.  In a similar fashion, the red dashed lines show the
cumulative distribution function of the CO(1-0) luminosity of the SPH particles,
but this time integrated from the other end of the $n$ axis.  For example, 50\%
of the luminosity results from particles whose density is larger than 100~\cmt. 
These particles constitute 9\% of the gas mass in that pixel.  The fits for PDR
models that do not consider mechanical heating are shown in
Figure-\ref{fig:paper3_fitting}.  We see that these models fit ratios involving
$J > 3-2$ transitions poorly compared to the mPDR fits, especially for the pixel
A and B which are closer to the center of the galaxy compared to pixels C and D.
In the remaining rows ({\bf B,C,D}) of Figure-\ref{fig:paper3_fitting}, we show
fits for pixels of increasing distances from the center of the galaxy.  We see
that as we move away further from the center,  the CO to \thco~ratios become
flat in general, close to the elemental abundance ratio of $^{13}$C/C which is
1/40 (see the green curve in the bottom row).  This is essential, because the
lines of both species are optically thin at the outer edge of the galaxy, hence
the emission is linearly proportional to the column density, which is related to
the mean abundance in the molecular zone.  Another observation is that the
distribution of the luminosity in a pixel becomes narrow at the edge of the
galaxy.  This is due to the low gas density and temperature in this region,
where there are less SPH particles whose density is close to the critical
density of the CO(1-0) line.  We also see that \gm~plays a minor role, where
both fits for a PDR with and without \gm~are equally significant.  The
parameters of the fits for the four representative pixels are listed in
Table-\ref{tbl:paper3_comaprison}.

\begin{table*}[!tbhp]
  \centering
  {\small
\begin{tabular}{c c | c c c c | c }
  \hline
  &  & $\log_{10}\left[\frac{n}{{\rm cm}^{-3}}\right]$ &   $\log_{10}\left[\frac{G}{G_0}\right]$    & $\log_{10}\left[\frac{\Gamma_{\rm mech}}{{\rm erg~cm}^{-3} {\rm~s}^{-1}}\right]$    & $\frac{A_V}{{\rm mag}}$ & $\chi^2_{\rm red}$ \\
  \hline
  disk & & & & & & \\ 
  galaxy & & & & & & \\
  \hline 
  {\bf A}& mPDR   & $1.8  \pm 0.1$ & $1.4 \pm 1.0$ & $-23.0 \pm 0.1$ & $22  \pm 2$   & $0.5 \pm 0.1$\\
  0.5 kpc& PDR    & $3.5  \pm 0.1$ & $4.7 \pm 0.3$ &       --        & $7.6 \pm 0.4$ & $3.7 \pm 0.1$\\
         & Actual & $1.63 \pm 0.1$ & $1.71\pm 0.1$ & $-23.1 \pm 0.1$ & $9.8 \pm 0.7$ &     --     \\
  \hline
  {\bf B}& mPDR   & $1.2  \pm 0.3$  & $0.6  \pm 0.9$ & $-23.5 \pm 0.3$ & $20.9 \pm 3.1$ & $0.7 \pm 0.2$ \\
  2 kpc  & PDR    & $3.5  \pm 0.1$  & $2.8  \pm 0.4$ &       --        & $5.1 \pm 0.6$  & $3.6 \pm 0.2$ \\
         & Actual & $1.16 \pm 0.2$ & $1.22 \pm 0.1$ & $-23.8 \pm 0.2$ & $4.7 \pm 0.7$  &     --        \\
  \hline
  {\bf C}& mPDR   & $1.5  \pm 0.3$  & $3.7  \pm 1.1$ & $-23.8 \pm 0.3$ & $10.1 \pm 2.4$ & $1.1 \pm 0.2$\\
  4 kpc  & PDR    & $2.7  \pm 0.3$  & $4.2  \pm 0.7$ &       --        & $6.6  \pm 1.2$ & $3.0 \pm 0.8$\\
         & Actual & $0.41 \pm 0.1$  & $0.53 \pm 0.1$ & $-24.9 \pm 0.2$ & $1.4 \pm  0.1$ &     --     \\
  \hline
  {\bf D}& mPDR   & $0.7   \pm 0.5$ & $1.6   \pm 1.3$ & $-26.5\pm 1.2$  & $3.6 \pm 1.7$ & $0.3 \pm 0.3$ \\
  7 kpc  & PDR    & $0.7   \pm 0.6$ & $3.3   \pm 0.9$ &       --        & $4.9 \pm 1.5$ & $0.6 \pm 0.4$ \\
         & Actual & $-0.23 \pm 0.1$ & $-0.18 \pm 0.1$ & $-26.6 \pm 0.2$ & $0.4 \pm 0.1$ &     --      \\
  \hline
  \hline
  dwarf  & & & & & & \\
  galaxy & & & & & & \\
  \hline 
         & mPDR   & $2.5 \pm 0.3$   & $3.4  \pm 0.8$  & $-23.4 \pm 0.3$  & $4.2 \pm 0.6$ & $0.6 \pm 0.4$\\
  0.5 kpc& PDR    & $2.5 \pm 0.3$   & $4.7  \pm 0.4$  &       --         & $5.4 \pm 0.4$ & $0.5 \pm 0.3$\\
         & Actual & $1.1 \pm 0.1$   & $1.33 \pm 0.1$  & $-24.9 \pm 0.2$  & $0.8 \pm 0.1$ &     --       \\
  \hline
\end{tabular}
  }
\caption{A comparison between the parameters fit using line ratios for the disk
and dwarf galaxy.
  The listed numbers are the averages of the fit parameters for pixels at a
  specific distance from the center, these distances are listed on the left of
  the Table.  For the dwarf galaxy we only show the fit parameters for the
  central 0.5 kpc.  For each pixel, we present the fit parameter (first row) for
  a mechanically heated PDR (mPDR), a pure PDR which does not consider
  mechanical feedback into its heating budget (second row),  and the actual  
  mean and standard deviation for these quantities for the SPH particles within
  the pixel of the luminosity map.  In the last column the statistic we minimize
  per degree of freedom, $\chi^2_{\rm red}$ is listed to compare the quality of
  the fits of the mPDR and the pure PDR models.
  \label{tbl:paper3_comaprison}}
\end{table*}

This fitting procedure can also be applied to the dwarf galaxy.  The main
difference in modeling the gas as a PDR in the disk and dwarf galaxy is that
the mechanical heating in the dwarf is lower compared to the disk, thus it does
not make a significant difference in the fits.  In
Figure-\ref{fig:paper3_line_ratio_maps_dwarf}, the line ratio maps show small
spatial variation, thus it is not surprising in having small standard deviations
in the fit parameters, in Table-\ref{tbl:paper3_comaprison} of the dwarf galaxy
compared to the disk galaxy. We note that the for regions at the outskirts of the
disk galaxy such as pixel {\bf C} and {\bf D}, we see significant deviations of the
recovered values of $n$, $G$, and $A_{\rm V}$ from the mPDR and PDR models. For the
mPDR models \gm the deviation could be due to degeneracies imposed by the strong dependence
of the emission on \gm in this relatively ``low'' density region compared to the central
region where \gm~ is the dominant heating mechanism.  As for the deviation of the best fit
PDR model for these pixels, they could
be linked to the fact that the quality of the fit is poor as indicated by the large $\chi^2$
values.  An advantage of using PDR models is that they do a better job at recovering the
actual value of $A_{\rm V}$ for pixels such as {\bf A} and {\bf B} ($R < 2$~kpc), but 
these fits are statistical less significant than the mPDR fits and they do not provide
information about the mechanical heating rate.

\section{Conclusion and discussion}

In this paper, we have presented a method to model molecular species emission
such as CO and \thco~for any simulation which includes gas and provides a local
mechanical heating rate and FUV-flux.
The method uses this local information to model the spatial dependence of the
chemical structure of PDR regions assumed to be present in the sub-structure of
the gas in the galaxy models.  These are then used to derive brightness maps
assuming the LVG approximation. The method for the
determination of the emission maps together with the ISM model
of the hydrodynamic galaxy models constitute a complete, self-consistent model
for the molecular emission from a star forming galaxies' ISM.  We compute the
emission of CO and \thco~ in rotational transitions up to $J=4-3$ for a model
disk-like and a dwarf galaxy. From the emission maps line ratios were
computed in order to constrain the physical parameters of the molecular gas using
one component PDR models.

We conclude the following:
\begin{enumerate}
\item Excitation temperature correlates positively with mechanical feedback in
equilibrium galaxies.
  This in turn increases for gas which is closer to the center of the galaxy. 
  The analysis presented in this paper allows estimates of mechanical feedback
  in galaxies which have high excitation temperatures in the center such as
  those observed by, e.g., \cite{muhle2007} and \cite{israel2009-1}.
\item Fitting line ratios of CO and \thco~ using a single mechanically heated
PDR component is sufficient
  to constrain the local \gm~throughout the galaxy within an order of magnitude,
  given the limitations of our modeling of the emission.  The density of the gas
  emitting in CO and \thco~ is better constrained to within half dex. This
  approach is not suitable in constraining the gas parameters of the dwarf
  galaxy, although the statistical quality of the fits is on average better than
  that of the disk galaxy.
\item Our approach fails in constraining the local FUV-flux.  It is
under-estimated by two
  orders of magnitude in mechanically heated PDR fits, and over-estimated by
  more than that in PDRs, which do not account for mechanical feedback.  This
  discrepancy is due considering one PDR component, where $ J < 3 - 2$ CO lines
  are optically thick and trace very shielded gas.  This discrepancy can also be
  explained by looking at the right column of Figure-\ref{fig:paper3_dist-app},
  where most of the emission results from SPH particles, which constitute a
  small fraction, and the rest of the gas is not captured by the PDR modeling.
\end{enumerate}

The main features of our PDR modeling with mechanical feedback is the higher
number of degrees of freedom it allows compared to LVG models and
multi-component PDR modeling.  We fit 10 line ratios while varying four
parameters in the case where \gm~was considered, and 3 parameters when \gm~was
not.  We note that the actual number of {\it independent} measurements of line
emission of CO and \thco~up to $J = 4-3$ is 8.  In LVG modeling the number of
free parameters is at least 5, since in fitting the line ratios we need to vary
$T$, $n({\rm H}_2$), $N$(CO), N(\thco) and the velocity gradient.  On the other
hand, PDR models assume elemental abundances and reaction rates. Moreover, the
uncertainties in the PDR modeling could be much larger and depend on the
micro-physics used in the PDR modeling \citep{vasyunin04, rolling07}. Thus, LVG
models have more free parameters compared to single PDR models used in our
method.  On the other hand, considering two PDR components increases the number
of free parameters to 9.  This renders the fitting problem over-determined
with more free parameters than independent measurements.  Although LVG models
are simple to run compared to PDR models,  they do not provide information about
the underlying physical phenomena exciting the line emission, which renders them
useless for constraining \gm.  Although a two component PDR model produces
better ``Xi-by-eye'' fits, these fits are statistically less significant than a
single PDR model, but physically relevant.  The main disadvantage of PDR
modeling is the amount of bookkeeping required to run these models and they are
computationally more demanding than LVG modeling\footnote{A typical PDR model
requires 10 time more CPU time than an LVG model in our simulations}, which
is why we resorted to using interpolation tables.

\subsection{Improvements and prospects}\label{subsec:improvementsandprospects}

Our models have two main limitations: 1) we make implicit assumptions on the
small scale structure of the galactic ISM and 2) we assume chemical equilibrium.
The first limitation can be improved by performing higher spatial resolution
simulations, which would result in gas that is better resolved and is denser
than the current maximum of $10^4$~\cmt in the disk galaxy.  Gas densities
$> 10^6$~\cmt allows us to consider transitions up to $J = 15 - 14$
which are more sensitive to \gm~compared to the $J=1-0$.
Galaxy scale simulations, which reach densities higher than $10^4$~\cmt~have
been performed by \cite{Narayanan2013-1} and required about $3 \times 10^7$
particles, and our method could be easily scaled to post-process such
simulations. In a follow up paper \citep{mvk16-2} we re-sampled the gas
density distribution of the disk simulation used in this paper in-order to
probe the effect of $n \gtrsim 10^4$~\cmt gas on the emission 
of $ 4-3 < J \le 15-14$ transitions of CO and \thco, in addition to $J \le 7-6$ 
transitions of HCN, HNC, and \hcop and their associated diagnostic line ratios
on galactic scales.

Having more data to fit helps in finding better diagnostics of \gm~ and narrower
constraints for it.  Another advantage in having emission maps for these high
$J$ transitions is having a higher number of degrees of freedom in fitting
for the emission line ratios. This renders the best fit PDR models statistically
 more significant.  Moreover,
it would be natural to consider multi-components PDR models in these fits, a
dense-component fitting the $J > 4 - 3$ transitions and low-density component
fitting the lower ones.  These components are not independent, low-$J$ emission
is also produced by the high density PDR components.  The high density component
PDR would generally have a filling factor 10 to 100 times lower than the low
density component, thus the high density model contributes mainly to the
high-$J$ transitions, whereas the low density models contributes mainly to the
low-$J$ emission and less to the high-$J$ emission.

In producing the synthetic line emission maps, we have used the LVG emission
from SPH particles, which assume PDR models as the governing sub-grid physics.
The main assumption in computing the emission using LVG models was the fixed
value of the micro-turbulence line-width $v_{\rm turb} = 1$~km/s for all the SPH
particles.
The peak of the distribution of $v_{\rm turb}$ of the SPH particles is located
at $\sim 3$~km/s. The optical depth in the lines are thus in reality smaller
than what we used in the calculations for the paper. When the lines become
optically thick it effectively reduces the critical density of those
transitions, and allows the excitation to higher energy states. The lines become
more easily optically thick, for normal cloud sizes ($A_{\rm V} = 5 - 10$) at
$v_{\rm turb} = 1$~km/s, which causes the peak of the CO ladder to be at a
higher rotational transition.
In order to quantify the shift of the peak of the CO ladder, we calculated a
grid with different turbulent velocities, $v_{\rm turb} = 1.0, 2.7, 5.0$,  and
10.0 km/s, representative of the turbulent velocities in the SPH simulation, and
used that grid to produce the resulting CO ladder.
We found that the peak of the CO ladder is located at CO(4-3) transition, when
computed from the distribution of $v_{turb}$, while the peak is located at
CO(6-5) for the calculations where only $v_{\rm turb} = 1.0$~km/s was used.
Although this causes a significant quantitative change in the emission, it does
not affect the general conclusions of the paper.
A possible consequence of considering $v_{\rm turb}$ to be constant is the 
fact that the \thco/CO line rations increase towards the center of our model
galaxies and are close to the elemental abundance ratio of $^{12}{\rm C}/^{13}{\rm C}$
$\sim 40$ in contrast to observations where this ratio ranges between 8 and 15 
\citep[e.g.][and references therein]{Buchbender13-01}.
The emission of CO are enhanced towards the center due to the increasing temperature of the gas.
Accounting for $v_{\rm turb}$ would most likely enhance the relative emission of CO compared
to \thco~where the computed line-ratios of the model galaxies would be in-line
with the observed trends and could also reduce $^{12}{\rm C}(1-0)/^{13}{\rm C}(1-0)$ to
the observed range of, which is currently a limitation to our 
proposed modelling approach.

An accurate treatment of the radiative transfer entails constructing the
synthetic maps by solving the 3D radiative transfer of the line emission using
tools such as LIME \citep{lime}.  This would eliminate any possible bias in the
fits.  Currently, we construct the emission maps by considering the mean flux of
the emission of SPH particles modeled as PDR within a pixel.  This results in a
distribution of the emission as a function of gas density as seen in
Figure-\ref{fig:paper3_dist} and Figure-\ref{fig:paper3_dist-app}.
When fitting the line ratios, we recover the mean physical parameters of the
molecular gas in that pixel associated with these emission.  Using tools such as
LIME, the micro-turbulence line width would be treated in a more realistic way
by using the actual local velocity dispersions in the small scale turbulent
structure of the gas, as opposed to adopting a fiducial 1 \kms as we have done
throughout the paper.  A major limitation of our sub-grid modelling is the 
assumption that the FUV flux at the surface of the PDR corresponds to that 
of the environment of the SPH particle.  The FUV flux and the FUV-driven heating
is strongly attenuated by the intervening dusty CNM, and the H$_2$ clouds
embedded in it. This is in contrast to turbulent and cosmic ray heating that
operate over large gas volumes.
An accurate treatment of the FUV radiative transfer could influence the modeled 
abundances of the molecular species and consequently the computed line emission fluxes.
that are used as diagnostics.

Not all H$_2$ gas is traced by CO, since the H/H$_2$ transition zone occurs
around A$_V = 0.1$~mag whereas the C/CO transition occurs around A$_V = 1$~mag.
When looking at Figure-\ref{fig:paper3_dist} and
Figure-\ref{fig:paper3_dist-app}, we see that most of the CO emission emanates
from a fraction of the gas.  The remaining ``major'' part of the gas can contain
molecular H$_2$, which is not traced by CO, but by other species such as C or
C$^+$, that do have high enough abundances to have bright emission
\citep{madden97-1, papadopoulos02-1, wolfire2010-1, pineda2014-1, israel15}.

Finally, we note that we have ignored any ``filling factor'' effects in
constructing the synthetic maps and the fitting procedure.  This is not relevant
to our method since in considering line ratios, and filling factors cancel out
in single PDR modeling of line ratios.
However, it is important and should be taken into account, when considering
multi-component PDR modeling and when fitting the absolute luminosities.

\begin{acknowledgements}
M.V.K would like to thank Francisco Salgado for advice on the fitting procedure
and Markus Schmalzl on producing the channel maps. M.V.K is grateful to Carla
Maria Coppola for some discussion and feedback. Finally, MVK would like to thank
the anonymous referee whose comments and suggestions helped
improve the paper significantly.
\end{acknowledgements}


\clearpage
\begin{appendix}{Appendix-A}
\label{app:a}
\setcounter{figure}{0}
\makeatletter
\renewcommand{\thefigure}{A\@arabic\c@figure} 

\begin{figure*}[!tbh]
   \center
   \includegraphics[scale=1.0, bb=54 284 558 507]{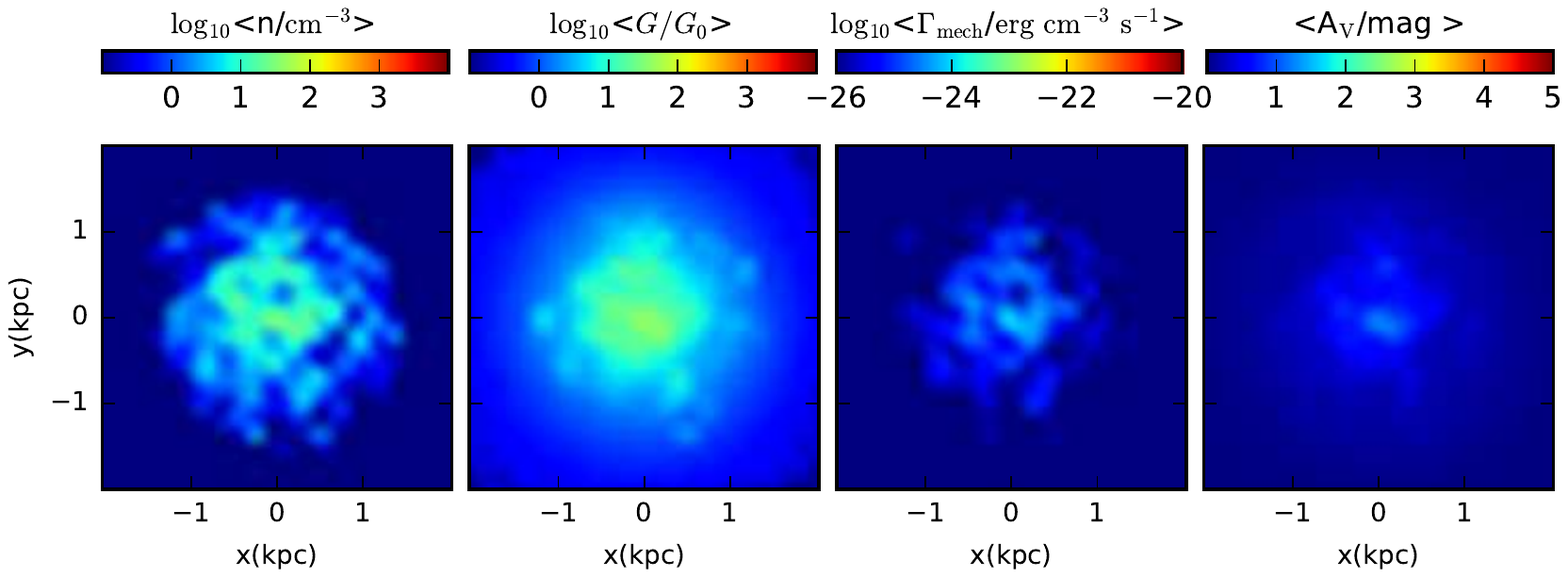}
   \caption{{\bf Left to right:} distribution maps of the gas
     density, FUV flux, mechanical heating rate and the $A_V$ of the model
     dwarf galaxy.
    \label{fig:paper3_n_g_gmech_av_maps_dwarf}} 
\end{figure*}

\begin{figure*}
  \center
  \includegraphics[scale=1.0, bb=0 0 357 433]{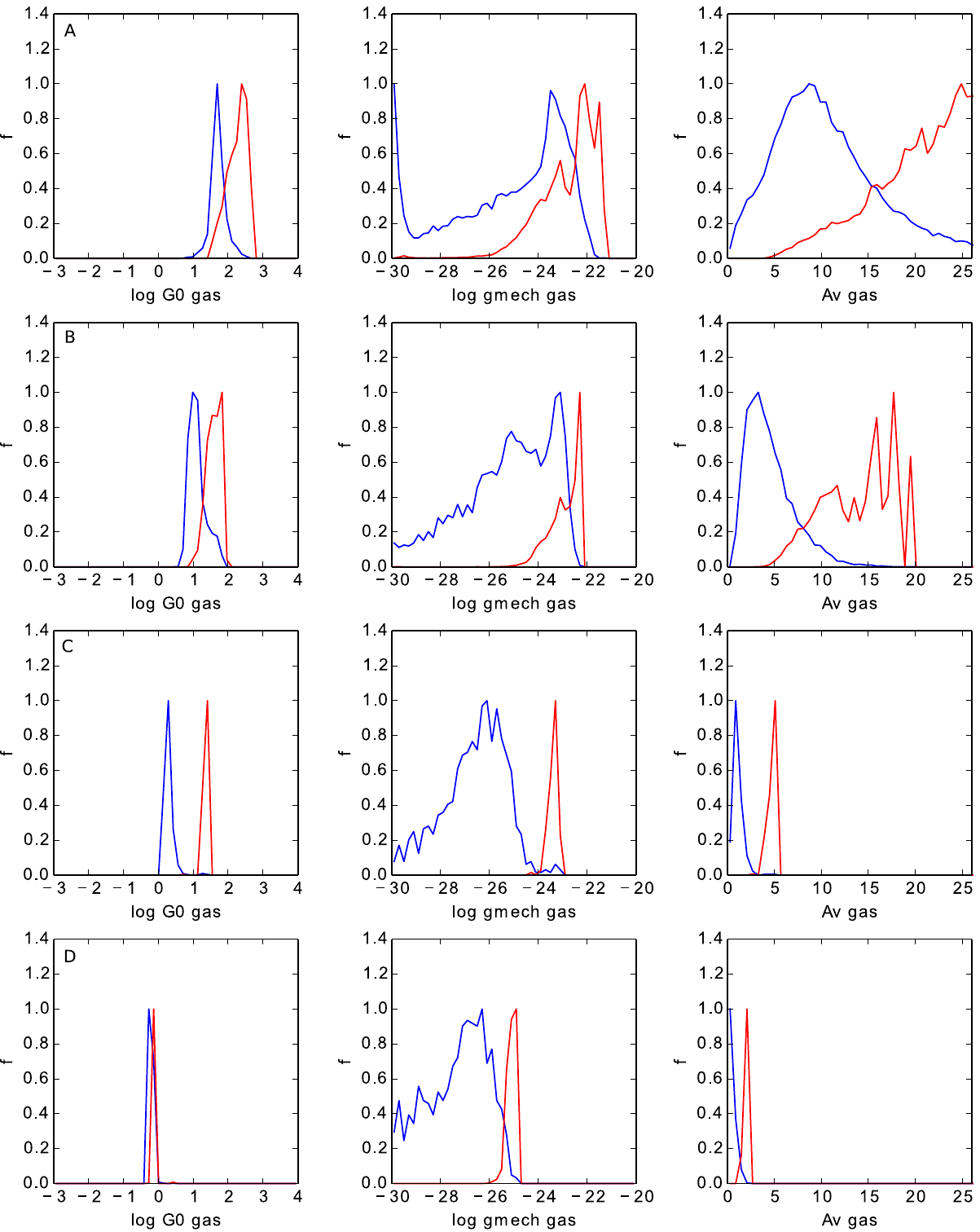}
  \caption{Distribution functions as a function of $G_0$, \gm~and $A_V$ for the
    representative pixels labeled A, B, C and D of
    Figure-\ref{fig:paper3_dist}.  The blue curves represent the distributions
    as a function of $G_0$ (first column), \gm~(second column) and $A_V$ (third
    column).  The red curves represent the fractional contribution of these
    distributions to the luminosity of their corresponding pixels.  Both curves
    are normalized to their peak so that we can visually compare the subset of
    these distributions which contribute to the luminosity.
\label{fig:paper3_dist-app}}
\end{figure*}

\begin{figure*}
  \center
  \includegraphics[scale=0.5, bb=0 0 718 742]{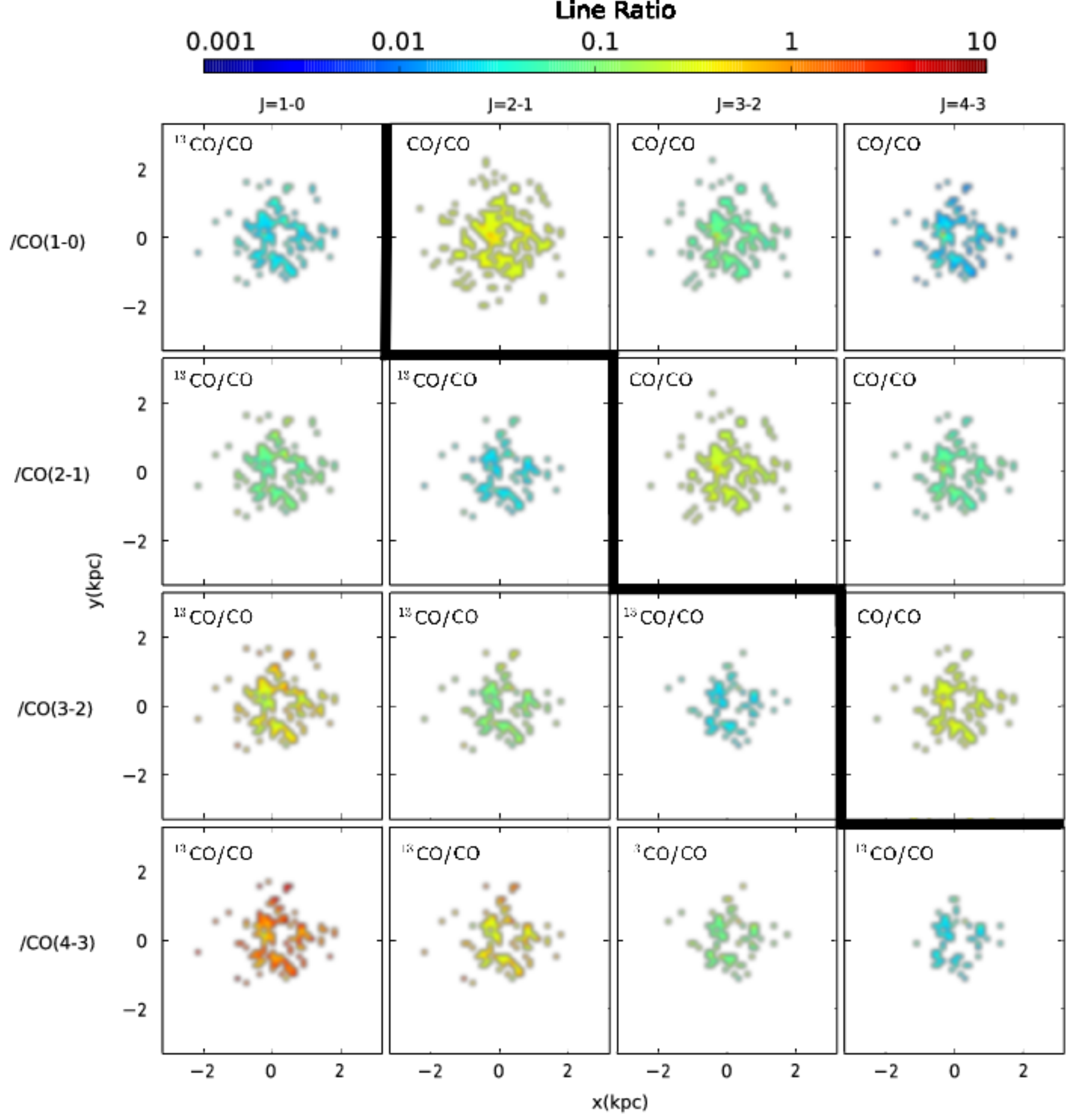}
  \caption{Line ratio maps for various transitions of CO and \thco~ for the
  dwarf galaxy.  The line
    ratios are generally spatially uniform and show slight variation, unlike the
    disk galaxy.
    These maps are also patchy, especially for line ratios involving $J = 3-2$
    and $J = 4-3$ transitions.  This is due to the low number of SPH particles
    with densities close to $10^4$~\cmt which are a pre-requisites for these
    transitions.  See caption of Figure-\ref{fig:paper3_line_ratio_maps} for
    more details on interpreting the maps.
    \label{fig:paper3_line_ratio_maps_dwarf}}
\end{figure*}

\end{appendix}

\end{document}